\documentclass[aps,prl,reprint,amsfonts,amssymb,amsmath]{revtex4-1}

\usepackage{hyperref}

\newcommand\cA{{\cal A}}

\newcommand\cJ{{\cal J}}

\newcommand\cN{{\cal N}}

\newcommand\cQ{{\cal Q}}

\newcommand{\eg}{\textit{e.g.}}

\newcommand{\ie}{\textit{i.e.}}

\newcommand{\nn}{\nonumber}
\newcommand{\fg}{\mathfrak{g}}

\DeclareMathOperator{\Tr}{Tr}

\begin{document}

\title{Exact two-dimensional superconformal R-symmetry and $c$-extremization}
\date{\today}

\author{Francesco Benini}
\author{Nikolay Bobev}
\affiliation{Simons Center for Geometry and Physics, Stony Brook University, Stony Brook, NY 11794-3636, USA}

\begin{abstract}
We uncover a general principle, dubbed $c$-extremization, which determines the exact R-symmetry of a two-dimensional unitary superconformal field theory with $\mathcal{N}=(0,2)$ supersymmetry. To illustrate its utility, we study superconformal theories obtained by twisted compactifications of four-dimensional $\cN=4$ super-Yang-Mills on Riemann surfaces, and construct their gravity duals.
\end{abstract}

\pacs{11.25.Tq, 11.25.Yb, 04.20.Jb}
\keywords{}

\maketitle

\textit{Introduction.}---Conformal field theories (CFTs) in two spacetime dimensions play a central r\^ole in describing critical phenomena and appear prominently in string theory. In two dimensions the conformal group is infinite-dimensional, and similarly other symmetries usually form infinite-dimensional algebras. As a consequence, such theories are tightly constrained and sometimes exactly solvable. Here we will be concerned with theories with at least $\cN=(0,2)$ superconformal symmetry:
they are interesting in their own right, play a pivotal r\^ole in type II and heterotic string compactifications \cite{Candelas:1985en, *Gepner:1987qi, *Banks:1987cy}, and are of some significance in mathematics because of their connection to vector bundles on Calabi-Yau manifolds \cite{Witten:1993yc}.

The $\cN=(0,2)$ superconformal algebra contains an Abelian right-moving (RM) Kac-Moody current $\omega$ called the R-symmetry current, under which the complex supercharge $\cQ$ is charged. This current
is important, as it determines the dimension of chiral primary operators and the Virasoro RM central charge $c_R$. In a non-conformal $\cN=(0,2)$ supersymmetric theory with an R-symmetry $U(1)_R$ and other Abelian flavor symmetries (under which $\cQ$ is not charged), the R-symmetry is not uniquely defined: mixing $U(1)_R$ with the flavor symmetries produces equally good R-symmetries. The R-current is in the same supermultiplet as the stress tensor $T_{\mu\nu}$, and mixing corresponds to improvement transformations of the multiplet \cite{Dumitrescu:2011iu}. On the contrary, if the theory flows to an infrared (IR) fixed point, the superconformal R-symmetry is singled out (improvement transformations are fixed by tracelessness of $T_{\mu\nu}$ and current conservation equations), and it is a non-trivial task to determine it.

In this letter we prove that in a unitary CFT (with mild normalizability assumptions) the superconformal R-symmetry is the linear combination of all IR Abelian symmetries that extremizes a quadratic function $c_R^\text{tr}$ of the coefficients. This function is entirely determined by the 't~Hooft anomalies of the theory, and thus it can be computed in the ultraviolet (UV), provided no accidental IR symmetry mixes in, even without a Lagrangian description of the theory. We call this principle $c$-extremization; it is analogous to $a$-maximization in four-dimensional superconformal field theories (SCFTs) \cite{Intriligator:2003jj}. The function $c_R^\text{tr}$ is dubbed the ``trial central charge'' because it is the would-be $c_R$ as a function of a trial R-symmetry. The central charge $c_R$ is a conformal anomaly, and supersymmetry relates it to the R-symmetry 't Hooft anomaly which is renormalization group (RG) invariant and easy to compute.

We illustrate the utility of $c$-extremization by computing the central charges of two-dimensional \mbox{$\cN=(0,2)$} SCFTs arising from the compactification of four-dimensional \mbox{$\cN=4$} super-Yang-Mills (SYM) on Riemann surfaces.
For gauge group $U(N)$, this corresponds to $N$ D3-branes wrapped on Riemann surfaces, and in the large $N$ limit we construct holographic dual AdS$_3$ solutions of type IIB supergravity: the holographic computation of the central charges perfectly agrees. More details and new examples will appear in \cite{BB}.

Such theories from genus-one Riemann surfaces were studied in \cite{Almuhairi:2011ws}, where the authors raised a puzzle about the computation of the central charges in field theory. We believe that the question is settled by $c$-extremization.

\textit{Anomalies.}---Local quantum field theories in two dimensions suffer from gauge and gravitational, but not mixed gauge-gravitational, anomalies \cite{AlvarezGaume:1983ig}. Consider a two-dimensional field theory where $U(1)^n$ is the Abelian part of the flavor symmetry group: then there are conserved currents
$J^I_\mu(x)$ with $I=1,\dots,n$, and a conserved stress tensor $T_{\mu\nu}(x)$. When the theory is coupled to non-dynamical (external) vector fields $A_\mu^I$ with field strengths $F_{\mu\nu}^I$, and to a curved background, the anomalous violation of current conservation is
\begin{equation}
\label{anomaly equations}
\nabla^\mu J^I_\mu = \sum_L \frac{k^{IL}}{8\pi}  F_{\mu\nu}^L \varepsilon^{\mu\nu} , ~~ \nabla_\mu T^\mu_\nu = \frac k{96\pi}  \varepsilon^{\alpha\rho} \partial_\alpha \partial_\beta \Gamma^\beta_{\nu\rho}
\end{equation}
where $\Gamma^\beta_{\nu\rho}$ are the Christoffel symbols and $\varepsilon^{\mu\nu}$ is the covariant antisymmetric tensor. The real constants $k^{IL}$ and $k$ are the 't~Hooft anomaly coefficients, and we have chosen a renormalization scheme in which $k^{IL}$ is a symmetric matrix, the stress tensor is symmetric, and local Lorentz transformations are non-anomalous. We emphasize that despite using the term ``gauge anomalies'', we always consider anomalies of global currents.

If the theory has a weakly coupled Lagrangian description, the coefficients $k^{IL}$ and $k$ get contribution from chiral fermions and bosons and are computed exactly by one-loop diagrams with two current insertions. \mbox{Spin-$\frac12$} (complex) Weyl fermions contribute as
\begin{equation}
\label{kanom}
k^{IL} = \Tr_\text{Weyl} \gamma^3 Q^I Q^L \;, \qquad\quad k = \Tr_\text{Weyl} \gamma^3 \;,
\end{equation}
where $\gamma^3$ is the chirality matrix that conventionally we take positive on right-movers, and $Q^I$ are the charge operators. Majorana-Weyl fermions contribute to $k$ as half of a Weyl fermion; real chiral bosons contribute to $k$ as Weyl fermions, and if linearly coupled to the vector fields as $\frac{Q^I}{\sqrt\pi} A^I_\mu \partial^\mu \phi$ they also contribute to $k^{IL}$ as Weyl fermions.
Regardless of the existence of a weakly coupled description, the anomaly coefficients $k^{IL}, k$ are well defined by the operator equations (\ref{anomaly equations}) and---as long as the symmetries are not broken---are invariant under RG flow \cite{tHooft:1980}.

If the theory is conformal, the anomaly coefficients $k^{IL},k$ are related to central terms in the conformal and current algebras (\ie{} in the operator product expansions (OPEs)) on flat space.
It is convenient to work in Euclidean signature ($x^0 = ix^0_E$) and in radial quantization, using holomorphic indices $z=x^1 + ix^0_E$, $\bar z=x^1 - i x^0_E$.  Following standard conventions (see \eg{} \cite{DiFrancesco:1997nk}) we define $T(z) = -2\pi T_{zz}(x)$, $\overline T(\bar z) = -2\pi T_{\bar z \bar z}(x)$, $j^I(z) = -i\pi J^I_z(x)$, $\bar\jmath^I(\bar z) = -i\pi J^{I}_{\bar z}(x)$.

We will consider CFTs in the following general class:
1) the CFT is unitary and the Virasoro generators $L_0, \overline L_0$ are bounded below;
2) the vacuum is normalizable.
Notable exceptions to the second condition are theories with non-compact free bosons. These assumptions lead to some standard properties:
First, in each conformal family there is a primary whose conformal dimensions $(\bar h, h)$ are non-negative. Second, an operator $\cA$ is holomorphic ($\bar\partial \cA = 0$) if and only if $\bar h = 0$, and it is anti-holomorphic ($\partial\cA = 0$) if and only if $h=0$ (the only $(0,0)$ operator is the identity). In particular conserved currents are either holomorphic (right-moving, RM) or anti-holomorphic (left-moving, LM).

Consider the conformal and current algebra OPEs:
\begin{eqnarray}
T(z) \, T(0) &\sim& \frac{c_R}{2z^4} + \frac{2T(0)}{z^2} + \frac{\partial T(0)}{z} \,,~~
j^I(z) \, j^J(0) \sim \frac{k_R^{IJ}}{z^2} \,, \nn \\
\overline T(\bar z) \, \overline T(0) &\sim& \frac{c_L}{2\bar z^4} + \frac{2\overline T(0)}{\bar z^2} + \frac{\bar\partial \overline T(0)}{\bar z} \,,~~\,
\bar\jmath^I(\bar z) \, \bar\jmath^J(0) \sim \frac{k_L^{IJ}}{\bar z^2} \,, \nn
\end{eqnarray}
where $\sim$ means equality up to regular terms.
Unitarity constrains $k_R^{IJ}$ and $k_L^{IJ}$ to be positive definite.
The OPEs between holomorphic and anti-holomorphic fields vanish.
We then have:
\begin{equation}
k^{IJ} = \begin{cases} k_R^{IJ} & \text{if $I,J$ are RM} \\ -k_L^{IJ} & \text{if $I,J$ are LM} \\ 0 & \text{otherwise} \;, \end{cases} \qquad\quad k = c_R - c_L \,.
\end{equation}

SCFTs with $\cN=(0,2)$ supersymmetry have two holomorphic \mbox{spin-$\frac32$} operators $T_F^\pm(z)$ (supercurrents) and a holomorphic \mbox{spin-1} operator $\omega(z)$ (R-symmetry current) in addition to the stress tensor $T(z), \overline T(\bar z)$.
The $\cN=2$ superconformal algebra is
\begin{eqnarray}
T(z) \, T(0) &\sim& \frac{c_R}{2z^4} + \frac{2T(0)}{z^2} + \frac{\partial T(0)}{z} ~, \nn\\
T(z) \, T_F^\pm(0) &\sim& \frac{3T_F^\pm(0)}{2z^2} + \frac{\partial T_F^\pm(0)}{z} ~, \label{N=2 SC algebra} \\
T(z) \, \omega(0) &\sim& \frac{\omega(0)}{z^2} + \frac{\partial \omega(0)}{z} ~, \quad~ T_F^{\pm}(z) \, T_F^{\pm}(0) \sim 0 \,, \nn\\
\omega(z) \, T_F^\pm(0) &\sim& \pm \frac{T_F^\pm(0)}{z}~, \qquad\quad~~~~~~ \omega(z) \, \omega(0) \sim \frac{c_R}{3z^2}, \nn\\
T_F^+(z) \, T_F^-(0) &\sim& \frac{2c_R}{3z^3} + \frac{2\omega(0)}{z^2} + \frac{2T(0)+ \partial\omega(0)}{z} ~. \nn
\end{eqnarray}
In particular the supercharge $\cQ$ has R-charge 1.
Equation (\ref{N=2 SC algebra}) fixes a relation between the central charge $c_R$ and the R-symmetry anomaly:
\begin{equation}
\label{central charge anomaly}
3k^{RR} = c_R \;.
\end{equation}

\textit{The superconformal R-symmetry.}---We wish to characterize the exact superconformal R-symmetry current $\Omega_\mu$ (where $\omega(z) = - i\pi \Omega_z$) in terms of anomalies, so that it is invariant under RG flow and independent of detailed knowledge of the physics at the IR fix point.
We consider a trial R-current $\Omega_\mu^\text{tr}$ constructed by mixing $\Omega_\mu$ with all flavor currents $J_\mu^I$:
\begin{equation}
\Omega_\mu^\text{tr}(t) = \Omega_\mu + \sum\nolimits_{I \, (\neq R)} t_I J_\mu^I \;.
\end{equation}
Then we construct a trial central charge $c_R^\text{tr}(t)$ proportional to the gauge anomaly of the trial R-symmetry:
\begin{equation}
c_R^\text{tr}(t) = 3\bigg( k^{RR} + 2 \sum_{I\, (\neq R)} t_I k^{RI} + \sum_{I,J\, (\neq R)} t_I t_J k^{IJ} \bigg) \;,
\end{equation}
which could be extracted from (\ref{anomaly equations}), from (\ref{kanom}) or from the two-point function $\langle \Omega_\mu(x) \, \Omega_\nu(0)\rangle$.

Superconformal symmetry imposes constraints on $k^{IJ}$. If $J_\mu^I$ is a LM flavor current then $k^{RI} = 0$, because $\Omega_\mu$ is RM. If $J_\mu^I$ is a RM flavor current, it is part of a supermultiplet. With $\cN=2$ superconformal symmetry, the multiplet of Kac-Moody currents $\cJ^A_+ = (\psi_{1,2}^A, j_{1,2}^A)$ is made of two $\cN=1$ current multiplets $(\psi_a^A, j^A_a)$ with $a=1,2$ \cite{Spindel:1988nh, *Kazama:1988uz} (the combined index $(A,a)$ runs over all RM flavor currents, covering a subset of the values of the index $I$). In superspace $\cJ^A_+$ is a holomorphic anti-chiral spinor superfield \cite{Hull:1989py}. In general there are constraints on the RM symmetry algebra (it has to admit a Manin triple \cite{Spindel:1988nh, *Kazama:1988uz, Parkhomenko:1992dq}), but in the Abelian case it only has to be even-dimensional.
The Abelian current algebra is described by the following non-vanishing OPEs:
\begin{equation}
\label{Abelian current algebra}
j_a^A(z) \, j_b^B(0) \sim \, \frac{\delta_{ab} \, k^{AB}}{z^2} \;, ~~ \psi_a^A(z) \, \psi_b^B(0) \sim  \, \frac{\delta_{ab} \, k^{AB}}z \;,
\end{equation}
where we have diagonalized the two-point function of the two currents in each $\cN=2$ multiplet, and the fermionic two-point function follows from Jacobi identities.
Unitarity requires $k^{AB}$ to be positive definite.

The action of the superconformal generators on the fields in the current multiplet is
\begin{eqnarray}
T(z) \, \psi^A_a(0) &\sim& \frac{\psi^A_a(0)}{2z^2} + \frac{\partial \psi^A_a(0)}z \;, \nn\\
T(z) \, j^A_a(0) &\sim& \frac{iq^A_a}{z^3} + \frac{j^A_a(0)}{z^2} + \frac{\partial j^A_a(0)}z \;, \label{OPEs SCCA} \\
T_F^\pm(z) \, \psi^A_a(0) &\sim& \frac{\delta_{ab} \mp i \varepsilon_{ab}}{\sqrt2} \, \Big( \frac{iq^A_b}{z^2} + \frac{j^A_b(0)}z \Big) \;, \nn\\
T_F^\pm(z) \, j^A_a(0) &\sim& \frac{\delta_{ab} \pm i \varepsilon_{ab}}{\sqrt2} \, \Big( \frac{\psi^A_b(0)}{z^2} + \frac{\partial\psi^A_b(0)}z \Big) \;, \nn\\
\omega(z) \, \psi^A_a(0) &\sim& i \varepsilon_{ab} \frac{\psi^A_b}z\;, \qquad \omega(z) \, j^A_a(0) \sim \varepsilon_{ab} \frac{q^A_b}{z^2}\;. \nn
\end{eqnarray}
The central terms $q^A_a$ (required to be real by unitarity) are called background charges \cite{Friedan:1985ge}: they are compatible with Jacobi identities and do not break superconformal symmetry.

Because of the central terms $q^A_a$ in the OPE $T(z)\, j^A_a(0)$ in \eqref{OPEs SCCA}, the currents $j^A_a$ are not primary operators. This also leads to a violation of current conservation on a gravitational background and of covariance on a gauge background. Since there are no mixed gauge-gravitational anomalies in two dimensions \cite{AlvarezGaume:1983ig}, one can cure the problem by adding local counterterms to the action, preserving superconformal symmetry.
From the point of view of the superconformal algebra, this corresponds to a redefinition of the triplet $(T, T_F^\pm, \omega)$ that preserves (\ref{N=2 SC algebra}) (up to a shift of the central charge $c_R$) but modifies (\ref{OPEs SCCA}) removing the background charges $q^A_a$. Most importantly,
the newly defined R-symmetry $\omega'$ has vanishing mixed gauge anomalies with all RM flavor currents $j^A_a$, as can be seen from the OPE $\omega'(z)\, j^A_a(0)$.

Let us show how to redefine the stress tensor multiplet. Under the linear shift
\begin{equation}
\begin{aligned}
T'(z) &= T(z) + i \alpha^A_a \, \partial j^A_a(z) \;, \\
T_F^{\pm\prime}(z) &= T_F^\pm(z) + \sqrt2\, i \alpha^A_a (\delta_{ab} \pm i \varepsilon_{ab}) \, \partial \psi^A_b(z) \;, \\
\omega'(z) &= \omega(z) + 2 \alpha^A_a \, \varepsilon_{ab} \, j^A_b(z) \;,
\end{aligned}
\end{equation}
the algebra (\ref{N=2 SC algebra}), (\ref{Abelian current algebra}), (\ref{OPEs SCCA}) is preserved up to the shifts:
\begin{equation}
\begin{aligned}
q^{\prime A}_a &= q^A_a - 2 k^{AB} \alpha^B_a \;,\\
c_R' &= c_R - 12 \alpha^A_a q^A_a + 12 \alpha^A_a \alpha^B_a k^{AB} \;.
\end{aligned}
\end{equation}
Since $k^{AB}$ is positive definite, it is always possible to cancel all central terms in (\ref{OPEs SCCA}) by taking $\alpha^A_a = \frac12 (k^{-1})^{AB} q^B_a$. The function $c'_R(\alpha)$ is quadratic with a positive definite second derivative; in fact it is minimized precisely at the value of $\alpha^A_a$ for which all RM currents are primaries. The central charge at that point, $c_R^0 = c_R - 3 q^A_a q^B_a (k^{-1})^{AB}$, is what is usually called \emph{the} central charge of the theory, and unitarity requires $c_R^0 > 0$. Moreover, at that point supersymmetry forbids mixed gauge anomalies between the superconformal R-current and RM flavor currents.

We have proven that at the IR fixed point, in a renormalization scheme in which there are no mixed violations
of covariance and flavor invariance, there are also no mixed anomalies between the superconformal R-current and flavor currents:
\begin{equation}
\label{orthogonality condition}
k^{RI} = 0 \qquad\qquad \forall\, I \neq R \;.
\end{equation}
Those conditions are RG invariant and can be imposed in the UV as well.
We can express (\ref{orthogonality condition}) as an extremality condition for $c_R^\text{tr}(t)$: $\Omega_\mu^\text{tr}(t_0)$ is the superconformal R-current for $t_0$ such that
\begin{equation}
\frac{\partial c_R^\text{tr}}{\partial t_I}(t_0)  = 0 \qquad\qquad \forall\, I \neq R \;.
\end{equation}
Since $c_R^\text{tr}(t)$ is a quadratic function, there is a unique solution. The function is actually maximized along directions $t_I$ that correspond to LM currents, and minimized along RM ones. This identifies which currents are LM or RM at the IR fixed point simply in terms of anomalies.
Notice that the assumption 2) discussed above might be relaxed if we are careful enough not to mix the R-symmetry with non-(anti)holomorphic currents.

We would like to point out the similarity between $c$-extremization and the minimization principle for the two-point function of the R-symmetry current in higher dimensions \cite{Barnes:2005bm}, as well as the maximization for $c_R$ in Landau-Ginzburg models observed in \cite{Melnikov:2009nh}.

\textit{An example.}---Let us study the $\cN=(0,2)$ two-dimensional theories arising at low energy from (twisted) compactifications of four-dimensional $\cN=4$ SYM on a closed Riemann surface $\Sigma_\fg$ of genus $\fg$. For gauge group $U(N)$, this corresponds to wrapping $N$ D3-branes on a holomorphic two-cycle $\Sigma_\fg$ in non-compact Calabi-Yau fourfolds.
To preserve some supersymmetry generically we have to twist the theory, \ie{} turn on a background gauge field $A_\mu$ coupled to the $SO(6)$ R-symmetry of the four-dimensional theory. We take the metric on the Riemann surface to be of constant curvature $\kappa$: $ds^2_\Sigma = e^{2h}(dx^2+dy^2)$, with $h = -\log \big( (1+x^2+y^2)/2 \big)$, $\kappa = 1$ for $\mathfrak{g}=0$; $h = \log(2\pi)/2$, $\kappa=0$ for $\fg=1$; and $h = -\log(y)$, $\kappa=-1$ for $\fg>1$. The background flux is then $F = dA = \sum_{I=1,2,3} F^I T^I$, with
\begin{equation}\label{fluxes}
F^I = - a_I \, e^{2h} \, dx\wedge dy \,, \qquad I=1,2,3\,,
\end{equation}
and $T^I$ the Abelian generators of the three factors in the $SO(2)^3$ Cartan subgroup of $SO(6)$ embedded block-diagonally.  The global symmetry of the IR theory is $SO(2)^3$, while the constants $a_I$ parametrize the twist and are quantized as $2(\fg - 1)a_I \in \mathbb{Z}$ for $\fg \neq 1$, and $a_I \in \mathbb{Z}$ for $\fg = 1$. To preserve supersymmetry we take
\begin{equation}
a_1+a_2+a_3 = -\kappa \;.
\end{equation}
For generic non-zero values of $a_I$, one has $\cN=(0,2)$ supersymmetry; if some of the $a_I$ vanish, supersymmetry is enhanced \cite{Maldacena:2000mw, Almuhairi:2011ws}. In what follows we will concentrate on the generic case. The trial R-symmetry is a linear combination of the generators of $SO(2)^3$:
\begin{equation}
T_R = \epsilon_1 T_1 + \epsilon_2 T_2 + (2-\epsilon_1-\epsilon_2) T_3 \;,
\end{equation}
where $\epsilon_{1,2}$ parametrize the mixing, and the R-charge of the complex supercharge $\cQ$ has been fixed to 1. To determine the correct superconformal R-symmetry
we can use $c$-extremization, and for that we need the two-dimensional anomalies.

The spectrum of massless chiral fermions of the two-dimensional theory is determined by the dimensional reduction of the gaugini of the four-dimensional theory \cite{Maldacena:2000mw}. The gaugini are in the representation $\textbf{2} \otimes \overline{\textbf{4}}$ of $SO(3,1)\times SO(6)$ and decompose under $SO(2)^3$ into chiral spinors of charges $A : (-\frac12,-\frac12,-\frac12)$, $B :(\frac12,\frac12,-\frac12)$, $C : (\frac12,-\frac12,\frac12)$, $D:(-\frac12,\frac12,\frac12)$. By the index theorem the number of RM minus LM fermions is
\begin{equation}
n_R^{(\sigma)} - n_L^{(\sigma)} = \frac1{2\pi} \int_{\Sigma}\Tr_\sigma F = - t_\sigma \, \eta_{\Sigma} \;,
\end{equation}
where $\sigma = \{A,B,C,D\}$, $\Tr_\sigma$ is taken in the representation $\sigma$, and $t_\sigma$ are the charges of the fermions in that representation under $A_\mu$, namely $t_A = \frac\kappa 2$, $t_B = \frac\kappa 2 + a_1 + a_2$, $t_C = -\frac\kappa 2 - a_2$, $t_D =  -\frac\kappa 2 - a_1$. We have also defined $\eta_{\Sigma} =\frac{1}{2\pi}\int_{\Sigma}e^{2h}dxdy$ with $\eta_{\Sigma}=|2(\mathfrak{g}-1)|$ for $\mathfrak{g}\neq 1$ and $\eta_{\Sigma}=1$ for $\mathfrak{g}=1$.
Taking into account that the fermions are in the adjoint representation of the gauge group, we can compute the trial central charge from \eqref{kanom} and \eqref{central charge anomaly}:
\begin{equation}
c_R^\text{tr}(\epsilon_i) = 3 \, d_G \sum\nolimits_\sigma \big( n^{(\sigma)}_R - n_{L}^{(\sigma)} \big) \big( q_R^{(\sigma)} \big)^2 \;,
\end{equation}
where $d_G$ is the dimension of the gauge group and $q_R^{(\sigma)}$ is the charge under $T_R$. One can extremize $c_R^\text{tr}(\epsilon_i)$ and find that at the critical point it takes the value
\begin{equation}
\label{cRexact}
c_R = \frac{12a_1a_2a_3 \eta_{\Sigma} d_G }{2(a_1a_2+a_1a_3+a_2a_3) - a_1^2 - a_2^2 - a_3^2}  \;.
\end{equation}
One also finds $c_R - c_L = \sum\nolimits_\sigma(n^{(\sigma)}_R - n_L^{(\sigma)}) = 0$ and thus there is no gravitational anomaly. If $c_R$ turns out to be non-positive, it means that one of our assumptions is not met.  We will see below that at large $N$, only in a particular infinite range of $a_I$ is there a unitary SCFT with normalizable vacuum.

For gauge group $SU(N)$ and at large $N$, one can find dual type IIB supergravity solutions.
We present here only the salient features of the solutions, deferring a detailed discussion to \cite{BB}. The metric is
\begin{equation}\begin{aligned}
\label{IIBsol}
ds^2_{10} &= \Delta^{1/2} \Big[ e^{2f} \, \frac{-dt^2 + dz^2 + dr^2}{r^2} + e^{2g} \, ds^2_\Sigma \Big] \\
 &+ \Delta^{-1/2}  \sum\nolimits_I (X^I)^{-1} \big( d\mu_I^2 + \mu_I^2 (d\varphi_I + A^I)^2 \big) \:,
\end{aligned}\end{equation}
and there is a non-trivial but not illuminating 5-form flux, that will be presented in \cite{BB}. We have defined
\begin{equation}\nn
\Delta = \sum\nolimits_I X^I \mu_{I}^2 \;,\quad\quad X^1X^2X^3 = 1 \;,\quad\quad \sum\nolimits_I \mu_I^2 = 1\;,
\end{equation}
while $A^I$ are the potentials for the fluxes $F^I$ in \eqref{fluxes}. The parameters specifying the solution are
\begin{eqnarray}
e^{2g} &=& \frac{a_1 X^2 + a_2 X^1}2 \;,~~~~
(X^1)^2 X^2 = \frac{a_1(a_2+a_3-a_1)}{a_3(a_1+a_2-a_3)} \;, \nn \\
e^f &=& \frac2{X^1+X^2+X^3} \;,~~
X^1 (X^2)^2 = \frac{a_2(a_1+a_3-a_2)}{a_3(a_1+a_2-a_3)} \;. \nn
\end{eqnarray}
For $\fg=0,1$ the solutions are regular and causal when two of the parameters obey $a_I>0$; for $\fg>1$ one needs two of the $a_I > 1/2$, or all three of them to be $a_I < 1/2$. In this range of $a_I$,
$\cN=4$ SYM flows to an IR fixed point and the AdS$_3$ supergravity solution \eqref{IIBsol} is the holographic dual to a normalizable ground state. The central charge at leading order in $N$ can be computed with standard holographic techniques:
\begin{equation}\label{ccsugra}
c_R = \frac{3R_\text{AdS$_3$}}{2G_{N}^{(3)}}= 6 \, e^{f+2g} \, \eta_{\Sigma} N^2 \;.
\end{equation}
Plugging $e^f$ and $e^{2g}$ in, one finds agreement with \eqref{cRexact} at leading order in $N$.

Further examples of $\cN=(0,2)$ SCFTs arising from M5-branes wrapped on four-manifolds will appear in \cite{BB}: the central charges will be computed by $c$-extremization from anomalies (extracted as in \cite{Benini:2009mz, *Alday:2009qq, *Bah:2012dg}), and matched by a holographic calculation.

\begin{acknowledgements}
We would like to thank C.~Beem, J.~Distler, C.~Herzog, K.~Jensen, I.~Melnikov, L.~Rastelli, M.~Ro\v cek, S.~Sethi, Y.~Tachikawa, B.~van~Rees and B.~Wecht for useful and informative discussions.
\end{acknowledgements}

\bibliography{c-ext-PRL}
\end{document}